\documentclass[twocolumn,showpacs,preprintnumbers,amsmath,amssymb]{revtex4}

\usepackage{graphicx}
\usepackage{dcolumn}
\usepackage{bm}

\begin{document}

\title{Electromagnetic Transition Strengths in Heavy Nuclei}

\author{B. Buck$^{\ast}$, A.C. Merchant$^{\ast}$ and S.M. Perez$^{\dagger\#}$}

\affiliation{%
$\ast$ Department of Physics, University of Oxford, Theoretical Physics, 
1 Keble Road, Oxford OX1 3NP, UK. \\
$\dagger$ 
Department of Physics, University of Cape Town, Private Bag, 
Rondebosch 7700, South Africa. \\
\# iThemba LABS, P.O.Box 722, Somerset West 7129, South Africa.
}%
\date{\today}
\begin{abstract}
\noindent We calculate reduced B(E2) and B(M1) electromagnetic transition
strengths within and between K-bands in support of a recently proposed
model for the structure of heavy nuclei. Previously, only spectra and a
rough indication of the largest B(E2) strengths were reported. The present more
detailed calculations should aid the experimental identification of
the predicted $0^+$, $1^+$ and $2^+$ bands and, in particular, 
act to confirm or refute the suggestion that the model $0^+$ and $2^+$ 
bands correspond to the well known and widespread beta and gamma bands. 
Furthermore they pinpoint transitions which can indicate the presence of a 
so far elusive $1^+$ band by feeding relatively strongly into or out
of it. Some of these transitions may already have been measured in 
$^{230}$Th, $^{232}$Th and $^{238}$U.

\end{abstract}
\maketitle

\section{Introduction} 
We have recently proposed a model for excited states
of heavy nuclei involving the coupling of a $2^+$ excitation to
a rotor \cite{[BMP_Spectrum]}. An important characteristic of this model
is the generation of excited $0^+$, $1^+$ and $2^+$ K-bands, at similar
excitation energies, as a generic feature. Such triplets of K-bands,
have been seen in the light nuclei $^{16}$0 \cite{[BBR_O16]}, 
$^{24}$Mg \cite{[BMH_Mg24]} and $^{40}$Ca \cite{[Ca40]},
and there is no obvious reason why they should not also
occur in heavier rare-Earth and Actinide nuclei.

It is difficult to make a conclusive identification of the 
three predicted K-bands from the calculated excitation energies alone. 
Their precise ordering depends on the details of the cluster-core
interaction and the excitation energy of the $2^+$ core state. 
If that interaction is overwhelmingly a quadrupole-quadrupole coupling, 
with model parameters appropriate to the Actinide region,
then our previous calculation \cite{[BMP_Spectrum]}
has the bands, in order of increasing excitation energy above 
the ground state band, as $K^{\pi}=1^+$, then $0^+$ and
finally $2^+$. However, even if we restrict ourselves to this
simplest form, we are unable to predict the absolute values of 
the bandhead excitation energies with any confidence because we
do not know the true strength of the non-central part of the 
cluster-core interaction.

Since $0^+$ beta and $2^+$ gamma bands are a generic feature
of the spectra of heavy nuclei, at about 1 MeV  above the 
ground state, we have chosen our interaction strength so as to place
the excited K-bands in this region.  The value required to do this
is not particularly large and indicates an intermediate strength,
rather than a truly strong coupling regime. Nevertheless, the 
suggestion that our calculated $0^+$ and $2^+$ bands might 
be identifiable with the beta and gamma bands needs to be 
confirmed or refuted.

An obvious form of confirmation would be that an accompanying 
$1^+$ band should be detected in the same region of excitation 
energies as the other two. Tell tale signs of such a band would 
include the discovery of a $1^+$ state (the bandhead) and also 
pairs of odd-J states (members of the proposed $1^+$ and $2^+$ bands) 
at similar excitation energies. However, as pointed out previously
\cite{[BMP_Spectrum]}, the population of the $1^+$ bandhead is
likely to be experimentally difficult. In addition, the 
intermediate strength interaction leads us to predict a
staggering of the energies of the states in the $1^+$ band,
which would make it unclear that they belong in a common band
from casual comparison of their excitation energies,
even if they were successfully populated. (This feature
of staggering is common to the $1^+$ bands of $^{16}$0 
\cite{[BBR_O16]}, $^{24}$Mg \cite{[BMH_Mg24]} and 
$^{40}$Ca \cite{[Ca40]},
as well as to the $1^-$ band of $^{238}$U
\cite{[U238a],[U238b]}.) Hence, additional information 
such as predicted electromagnetic transition strengths
within and between the proposed band members is needed to
enable experimental groups to recognise the states of
the $1^+$ band if they do succeed in exciting them.

Here, we seek to improve on our previous rough indications 
of where strong E2 transitions should be expected
by calculating in-band and cross-band reduced B(E2) and
B(M1) transition strengths for the ground and excited
$0^+$, $1^+$ and $2^+$ K-bands in far greater detail.
We cannot provide a definitive and unambiguous account
of the E2 and M1 transitions because not only does their
direction depend on the details of the ordering of the 
states in the spectrum, but also the wave functions 
are sensitive to how close in energy the states they 
represent lie. Subject to these provisos, this paper
presents a detailed account of the calculated E2 and M1 
transition strengths appropriate to the states generated
previously \cite{[BMP_Spectrum]} which should be of 
considerable assistance to experimental groups searching 
for the predicted triplet of K-bands.

Theoretical studies of positive parity bands in the Actinide 
nuclei including calculation of some electromagnetic transition
rates (but generally only in-band rather than cross-band) have
also been presented within the cranked RPA 
\cite{[Neergard],[Vogel],[Ward],[Kuliev]},
the collective model \cite{[Minkov1],[Minkov2],[Raduta]}, 
the interacting boson approximation
\cite{[Zamfir],[Cottle]}, the variable moment of inertia model 
\cite{[Lenis]} and the alpha particle cluster model \cite{[Schneidman]}.

In the next Section we briefly outline the structure
model leading to the generation of K-bands. Then, we
calculate reduced B(E2) and B(M1) electromagnetic transition
strengths within and between these K-bands. Insofar as 
possible, we compare our calculated B(E2) strengths with
experimental values for some isotopes of Th and U.
Finally we summarize our conclusions.

\section{Cluster model to generate K-bands}

We model a nucleus ($Z$,$A$) as a core ($Z_1$,$A_1$) and a 
cluster ($Z_2$,$A_2$) interacting via a deep, local 
potential $V(R)$ consisting of nuclear and Coulomb
terms $V_N(R) $ and $V_C(R)$, respectively, where $R$ is the
separation of their centres. We parametrise the
nuclear term as \cite{[univ]}:
\begin{eqnarray}
V_N(R) &=& -V_0 \Biggl \{ {x \over [1 + \exp{((R-R_0)/a))} ] } 
\nonumber \\\ &+&
{1-x \over [1+\exp{((R-R_0)/3a)} ]^3 } \Biggr \}
\end{eqnarray}
and take $V_C(R)$ to represent a cluster point charge 
interacting with a uniformly charged spherical core of radius $R_0$.
If the cluster and core were both restricted to their 
$0^+$ ground states a single band of states would be 
produced by solving the Schr\"odinger equation with
this potential for a fixed value of the
global quantum number $G=2n+L$ ($n$ the number of internal 
nodes in the radial wave function and $L$ the orbital angular momentum).
The value of $G$ must be chosen large enough to satisfy the major 
requirements of the Pauli exclusion principle by 
excluding the cluster nucleons from states occupied by the 
core nucleons. This can be achieved in the Actinide region
by taking $G=5A_2$. This programme leads to a band of states 
$L^{\pi} = 0^+, 2^+, 4^+ \ldots G^+$ and excitation energies $E_L$.

The situation becomes a little more complicated if we accept
the possibility of the core being in either its ground state
or an excited state having spin-parity $I^{\pi}$
(we restrict attention to $2^+$ excitations here) and 
excitation energy $\epsilon$. The cluster-core
potential may now contain non-central terms and the
system must be described in terms of coupled basis states
$\vert (IL) JM \rangle$ formed by combining the core spin $I$
with the relative orbital angular momentum $L$ to obtain a
total angular momentum $J$. The simplest form of non-central
potential compatible with considerations of time reversal and
parity invariance is a quadrupole-quadrupole interaction.
The matrix elements of such a non-central potential between 
our coupled basis states are
\begin{eqnarray}
V_{LL^{\prime}II^{\prime}}^J = i^{L^{\prime}-L+I^{\prime}-I}(-1)^{J+L+L^{\prime}} \beta \hat{I}\hat{I^{\prime}}\hat{L}\hat{L^{\prime}}
\nonumber \\
\langle L0L^{\prime}0 \vert 20 \rangle \langle I0I^{\prime}0 \vert 20 \rangle
W(L^{\prime}I^{\prime}LI;J2)
\end{eqnarray} 
where $\hat{L}=\sqrt{(2L+1)}$ etc.
By combining these with the diagonal elements of the
cluster-core rotational motion, simplified to
$\alpha L(L+1)$, and the core excitation energy, zero
for $I=0$ and $\epsilon$ for $I=2$, and assuming that
all the radial integrals can be parametrised by a single
strength $\beta$ we can obtain eigenvalues and eigenvectors
by diagonalization of low-dimensional matrices. Figure 1
shows the spectrum resulting from such a calculation 
using the parameters of Ref.\cite{[BMP_Spectrum]}, where 
more details of the calculation are given.

\section{Calculation of electromagnetic transition strengths}

We calculate reduced B(E2) and B(M1) electromagnetic transition
strengths appropriate to the spectrum shown in Fig.1.
The general features of these results are expected to be 
common to all those nuclei for which the triplet of K-bands
produced by coupling a $2^+$ excitation to a rotor is an 
appropriate description of part of the spectrum. There
should be strong in-band E2 transitions (typically a few hundred
Weisskopf units in the Actinide region), related to one
another rather closely by Clebsch-Gordan coefficients, 
and usually much weaker cross-band E2 transitions of typically
a few Weisskopf units (although there turn out to be a few
interesting exceptions to this general statement). 

Magnetic dipole transitions  are possible between some of the
levels due to mixing of the relative motion $L$-values
(e.g. the $3^+$ states are a mixture of $L=2, 4 \otimes I=2$
while the $2^+$ states  contain mixtures of $L=0, 2, 4 \otimes I=2$ 
as well as $L=2 \otimes I=0$) which allow a transition
to take place between certain components of these states
without changing $L$ or $I$. These B(M1) strengths should be
rather small, typically 0.01 Weisskopf units, but the
simple existence of such transitions between $0^+$ and 
$1^+$ states, where E2 transitions are impossible, can provide 
a strong indication of our proposed band structure.

The precise details of the transition rates will vary 
from nucleus to nucleus. Accidental near-degeneracy of excited
states in a particular nucleus can lead to exceptionally large 
mixing with accompanying strong E2 transition strengths. 
It must also be borne in mind that any given real nucleus may have 
some of the predicted levels in a different order from those 
exhibited in Fig.1 so that the transitions go in the opposite direction.
Indeed, the excited bandhead ordering may differ from that shown in
Fig.1. Even then, our results can still be useful because it is 
straightforward to transform the calculated strength for $J_i \to J_f$ into
a corresponding value for $J_f \to J_i$ on multiplication by
$(2J_i+1)/(2J_f+1)$.

To maintain generality as far as possible we give the 
calculated B(E2) transition strengths as ratios to the value
for B(E2; $2^+ \to 0^+$) in the ground state band. However, 
we give the B(M1) strength estimates directly in Weisskopf units
because they only depend on the Z/A ratios (which are all very 
similar in the Actinide region) with the appropriate Weisskopf 
unit independent of the charge and mass of the nucleus in
question.

\subsection{Reduced B(E2) strengths}

In the limit of strong coupling there would be no cross-band
E2 transitions at all, and the in-band transitions would be
given by (see for example Ref.\cite{[Leander]})
\begin{equation}
B(E2; J_i \to J_f) = {5\over 4\pi} Q_2^2 
\langle J_iK 20 \vert J_fK \rangle^2
\end{equation}
where $Q_2$ is the constant intrinsic electric quadrupole 
transition strength. We have checked that by increasing 
the strength of our quadrupole-quadrupole interaction 
to very large values this situation does indeed emerge 
from our calculation. Even with a much reduced interaction
strength, Eq.(3) still provides a reasonable zeroth order 
approximation to the calculated in-band reduced E2 transition 
strengths, as can be seen from Table 1. The stretched E2
transitions between states with $J$ differing by 2 are
uniformly strong and much more prominent than most of the 
transitions between states whose $J$ values differ by 1. 
This simple distinction does not hold at low spins where
many states lie close together in energy and consequently 
their wave functions are thoroughly mixed. There we see
equally strong $1^+ \to 2^+$ and $3^+ \to 2^+$ transitions
within the $K^{\pi}=1^+$ band and $3^+ \to 2^+$, $4^+ \to 3^+$  
and $5^+ \to 4^+$ transitions in the  $K^{\pi}=2^+$ band. 
The general pattern  of strong $J \to J-2$ and much weaker
$J \to J \pm 1$ transitions reasserts itself at higher values of
$J$ as state mixing gradually decreases and individual states
become more widely separated in energy.

In view of the staggering of energy levels predicted for the 
$K^{\pi}=1^+$ band, this general preference for stretched E2 transitions
could easily lead to the perception that the $K^{\pi}=1^+$
band consists of two separate
bands with angular momentum sequences $2^+, 4^+, 6^+, \ldots$
and $1^+, 3^+, 5^+, \ldots$ such that their common origin is
not apparent. See also Fig.3 of Ref.\cite{[BMP_Spectrum]}.

\begin{center}
\begin{table}[htb]
\caption{Calculated in-band reduced E2 strengths B(E2; $J_i \to J_f$)
for $K^{\pi} = 0^+_{\rm gs}, 1^+, 0^+_{\rm ex}$ and $2^+$ bands
in heavy nuclei. Values are given as ratios relative to the strength
of the $2^+ \to 0^+$ transition in the ground state band.
See text for details.}

\begin{tabular}{lcccc}
\hline
& & & & \\
Transition & \multispan 4 $K^{\pi}$  \\
& $0^+_{\rm gs}$  & $1^+$ & $0^+_{\rm ex}$ & $2^+$ \\
& & & & \\
\hline
$2^+ \to 0^+$ & 1.00 & & 0.60 & \\
$4^+ \to 2^+$ & 1.43 & 0.93 & 1.06 & 0.38 \\
$6^+ \to 4^+$ & 1.58 & 1.09 & 1.22 & 0.74 \\
$8^+ \to 6^+$ & 1.67 & 1.14 & 1.30 & 0.92 \\
$10^+ \to 8^+$ & 1.72 & 1.15 & 1.34 & 1.03 \\
& & & & \\
\hline
& & & & \\
Transition &  $K^{\pi}$ & Transition & $K^{\pi}$  & \\
& $1^+$ & & $2^+$ & \\
& & & & \\
\hline
& & & & \\
$1^+ \to 2^+$ & 1.45 & $3^+ \to 2^+$ & 1.21 &  \\
$3^+ \to 4^+$ & 0.22 & $4^+ \to 3^+$ & 0.83 & \\
$3^+ \to 1^+$ & 0.80 & $5^+ \to 4^+$ & 0.70 & \\
$3^+ \to 2^+$ & 0.63 & $5^+ \to 3^+$ & 0.64 & \\
$5^+ \to 6^+$ & 0.07 & $6^+ \to 5^+$ & 0.37 & \\
$5^+ \to 3^+$ & 1.25 & $7^+ \to 6^+$ & 0.42 & \\
$5^+ \to 4^+$ & 0.27 & $7^+ \to 5^+$ & 0.89 & \\
$7^+ \to 8^+$ & 0.02 & $8^+ \to 7^+$  & 0.18 & \\
$7^+ \to 5^+$ & 1.39 & $9^+ \to 8^+$ & 0.29 & \\
$7^+ \to 6^+$ & 0.17 & $9^+ \to 7^+$ & 1.03 & \\
$9^+ \to 10^+$ & 0.01 & $10^+ \to 9^+$ & 0.10 & \\
$9^+ \to 7^+$ & 1.45 & & & \\
$9^+ \to 8^+$ & 0.12 & & & \\
& & & & \\
\hline
\end{tabular}
\end{table}
\end{center}
In general, the E2 transitions in our model are produced by
a sum of three terms. One is due to intrinsic excitations of the
core, another is due to intrinsic excitations of the cluster
and the third is due to the relative motion of cluster about
core. Since we restrict the cluster to its $0^+$ ground state
there is no contribution from intrinsic cluster excitations
and the electromagnetic operator $M(E2)$ reduces to
\begin{equation}
M(E2) = a_{\rm core}r^2Y_2(\hat{\bf r}) + a_{\rm rel}
R^2Y_2(\hat{\bf R})
\end{equation}
where $r$ is a core internal coordinate and $R$ the cluster core
separation, and $a_{\rm core}$ and $a_{\rm rel}$ are the relevant
charge factors, with $a_{\rm rel} = {Z_1A_2^2 + Z_2A_1^2 \over (A_1+A_2)^2 }$.
We follow the convention of Brink and Satchler \cite{[BS]}
in defining reduced matrix elements of rank $k$ tensor operators
through the Wigner-Eckart theorem by
\begin{equation}
\langle J_f M_f \vert T_k^q \vert J_i M_i \rangle = 
(-1)^{2k} \langle J_f M_f \vert J_i M_i k q \rangle
\langle J_f \vert\vert T_k \vert\vert J_i \rangle
\end{equation}
so that we have to modify Eq.(3C-17) of Ref.\cite{[BM_E2]}
and write the reduced transition strengths as
\begin{equation}
B(E2; J_i \to J_f) = {2J_f+1 \over 2J_i+1} 
\vert \langle J_f \vert\vert M(E2) \vert\vert J_i \rangle \vert^2
\end{equation}
and the corresponding Weisskopf unit is \cite{[BM_WU]}
\begin{equation}
B(E2)_{\rm Wu} = {0.747 \over 4\pi}(A_1+A_2)^{4/3} \ \ \ {\rm e}^2 \ {\rm fm}^4
\end{equation}
To keep our results as general as possible we need to avoid
explicit dependence on charge and mass values. To this end 
we note that in $^{208}$Pb, a typical core for our model, 
the B(E2 $\downarrow$) from the excited
$2^+$ state at 4.08524 MeV to the $0^+$ ground state is measured 
as $8.4 \pm 0.5$ Wu. The strengths of the corresponding E2 transitions 
in our parent nuclei are of order 200--300 Wu (which we model as a combination of
intrinsic core and relative motion contributions). We use this
information to estimate the relative strengths of the orbital 
and core excitation terms in the $M(E2)$ operator. This is
sufficient because we are not concerned with the absolute values
of the B(E2) strengths, and intend to calculate them only as 
ratios to the strength of the ground state band $2^+ \to 0^+$
transition. If we write $M(E2) = a_{\rm rel}R^2Y_2(\hat{\bf R}) +
a_{\rm core}r^2Y_2(\hat{\bf r})$, then we can estimate 
$a_{\rm rel}/a_{\rm core}$ by fitting to a ratio of a mid mass
range Actinide B(E2) value and the $^{208}$Pb B(E2) value which, 
on ignoring the niceties of angular momentum coupling, yields
\begin{equation}
{(a_{\rm rel} + a_{\rm core})^2 \over a_{\rm core}^2 } \approx {250 \ 
  {\rm W.u.} \over 8.4 \ {\rm W.u.}}
\end{equation}
We therefore expect relative coefficients of the corresponding terms
in the M(E2) matrix elements to be roughly in the ratio
$[\sqrt{(250/8.4)}-1] : 1$, which is to say roughly between
4:1 and 5:1. In view of this, we use an $M(E2)$ operator 
\begin{equation}
M(E2) = 0.25 r^2Y_2(\hat{\bf r}) + R^2Y_2( \hat{\bf R})
\end{equation}
and present our results in Tables 1 and 2 as ratios to the strength
of the $2^+ \to 0^+$ transition in the ground state band.

Table 1 shows our results for in-band E2 transitions.
Stretched E2 transitions are all strong and due principally
to the relative motion term in the $M(E2)$ operator.
Their values are not generally far from the expectations
of the rotational model given in Eq.(3), despite the
strength of our quadrupole-quadrupole interaction being 
so weak that only intermediate strength coupling is produced. 
Therefore, the measurement of in-band stretched E2 transitions will 
not distinguish between the two models to any convincing degree.
The weaker values between states with $J$ differing by less than 2
can largely be attributed to angular momentum coupling effects 
(i.e. Clebsch-Gordan coefficients).

On the other hand, our predictions for the cross-band E2,
transitions shown in Table 2,
are very different from the strong-coupling rotational model
according to which they should all be zero. It is interesting to
note that if we had only included the relative motion term
in the $M(E2)$ operator, we would have obtained very weak
cross-band transitions and so reached an indistinguishable
conclusion, despite not having attained a truly strong
coupling limit. However, the presence of the core excitation 
term in $M(E2)$ gives rise to small, but nevertheless
significant cross-band E2 transition strengths --- a few hundreths
of those seen for the in-band cases. These strengths typically amount to a  
few Wu and should be experimentally measurable. We also note that
in a few cases, where the state mixing is especially favourable,
there are some transitions between the excited $0^+$ and $1^+$
bands whose strengths are about one tenth (or even larger) than 
characteristic in-band values. These cross-band features 
differ substantially from the
expectations of the strong-coupling rotational model and 
thus distinguish our model from it. At present, the experimental
data in the Actinide region (which we examine later) are
too sparse to draw any firm conclusion. However, similar 
effects are expected in the rare-Earth region, and we hope 
to turn our attention to some of those nuclei in due course.
\begin{widetext}
\begin{center}
\begin{table}[htb]
\caption{Calculated cross-band reduced E2 strengths B(E2; $J_i \to J_f$)
between $K^{\pi} = 0^+_{\rm gs}, 1^+, 0^+_{\rm ex}$ and $2^+$ bands
in heavy nuclei. Values are given as ratios relative to 
the strength of the $2^+ \to 0^+$ transition in the ground state band.
See text for details.}

\begin{tabular}{lrrrrrrr}
\hline
& & & & & & & \\
Transition & $1^+ \to  0^+_{\rm gs}$  & 
$0^+_{\rm ex} \to 0^+_{\rm gs}$  & 
$0^+_{\rm ex} \to 1^+$  & 
$ 2^+ \to 0^+_{\rm gs}$  & 
$ 2^+ \to 1^+$  & 
$ 2^+ \to  0^+_{\rm ex}$ &
$ 0^+_{\rm ex} \to 2^+$ \\
& & & & & & & \\
\hline
$0^+ \to 2^+$ &        & 0.0533 & 0.6878 &        &        &        & \\
$2^+ \to 0^+$ & 0.0299 & 0.0002 &        & 0.0076 &        & 0.0166 & \\
$1^+ \to 2^+$ & 0.0479 &        &        &        &        &        & \\
$2^+ \to 1^+$ &        &        & 0.2301 &        & 0.0019 &        & \\
$2^+ \to 2^+$ & 0.0143 & 0.0060 & 0.0184 & 0.0154 & 0.0051 & 0.0841 & \\
$2^+ \to 3^+$ &        &        & 0.1898 &        & 0.0000 &        & \\
$2^+ \to 4^+$ & 0.0022 & 0.0471 & 0.0070 & 0.0027 & 0.0029 & 0.0129 & \\
$3^+ \to 1^+$ &        &        &        &        & 0.0032 &        & \\
$3^+ \to 2^+$ & 0.0179 &        &        & 0.0125 & 0.0051 & 0.0253 & \\
$3^+ \to 3^+$ &        &        &        &        & 0.0007 &        & \\
$3^+ \to 4^+$ & 0.0288 &        &        & 0.0132 & 0.0116 & 0.0975 & \\
$3^+ \to 5^+$ &        &        &        &        & 0.0021 &        & \\
$4^+ \to 2^+$ & 0.0354 & 0.0000 & 0.0010 & 0.0026 & 0.0054 & 0.0045 & \\
$4^+ \to 3^+$ &        &        & 0.0776 &        & 0.0001 &        & \\
$4^+ \to 4^+$ & 0.0065 & 0.0118 & 0.0013 & 0.0140 & 0.0084 & 0.0934 & \\
$4^+ \to 5^+$ &        &        & 0.0737 &        & 0.0017 &        & \\
$4^+ \to 6^+$ & 0.0010 & 0.0399 & 0.0001 & 0.0108 & 0.0108 &        & \\
$5^+ \to 3^+$ &        &        &        &        & 0.0038 &        & \\
$5^+ \to 4^+$ & 0.0232 &        &        & 0.0075 & 0.0047 & 0.0012 & \\
$5^+ \to 5^+$ &        &        &        &        & 0.0002 &        & \\
$5^+ \to 6^+$ & 0.0217 &        &        & 0.0185 & 0.0183 & 0.1461 & \\
$5^+ \to 7^+$ &        &        &        &        & 0.0071 &        & \\
$6^+ \to 4^+$ & 0.0334 & 0.0001 & 0.0000 & 0.0012 & 0.0026 & 0.0240 & 0.0156 \\
$6^+ \to 5^+$ &        &        & 0.0389 &        & 0.0001 &        & \\
$6^+ \to 6^+$ & 0.0045 & 0.0180 & 0.0081 & 0.0085 & 0.0059 & 0.0683 & \\
$6^+ \to 7^+$ &        &        & 0.0295 &        & 0.0039 &        & \\
$6^+ \to 8^+$ & 0.0007 & 0.0292 & 0.0001 & 0.0212 & 0.0192 &        & \\
$7^+ \to 5^+$ &        &        &        &        & 0.0029 &        & \\
$7^+ \to 6^+$ & 0.0250 &        &        & 0.0047 & 0.0032 & 0.0031 & \\
$7^+ \to 7^+$ &        &        &        &        & 0.0001 &        & \\
$7^+ \to 8^+$ & 0.0178 &        &        & 0.0213 & 0.0220 &        & \\
$7^+ \to 9^+$ &        &        &        &        & 0.0131 &        & \\
$8^+ \to 6^+$ & 0.0293 & 0.0001 & 0.0002 & 0.0006 & 0.0012 & 0.0216 & 0.0057 \\
$8^+ \to 7^+$ &        &        & 0.0242 &        & 0.0001 &        & 0.1330 \\
$8^+ \to 8^+$ & 0.0035 & 0.0217 & 0.0134 & 0.0045 & 0.0033 & 0.0381 & \\
$8^+ \to 9^+$ &        &        & 0.0115 &        & 0.0042 &        & \\
$8^+ \to 10^+$& 0.0006 & 0.0206 & 0.0000 & 0.0288 & 0.0261 &        & \\
$9^+ \to 7^+$ &        &        &        &        & 0.0020 &        & \\
$9^+ \to 8^+$ & 0.0249 &        &        & 0.0031 & 0.0022 & 0.0097 & \\
$9^+ \to 9^+$ &        &        &        &        & 0.0000 &        & \\
$9^+ \to 10^+$& 0.0156 &        &        & 0.0223 & 0.0246 &        & \\
$10^+ \to 8^+$& 0.0241 & 0.0001 & 0.0003 & 0.0003 & 0.0006 & 0.0106 & 0.0006 \\
$10^+ \to 9^+$ &       &        & 0.0170 &        & 0.0001 &        & 0.1219 \\
$10^+ \to 10^+$& 0.0029& 0.0228 & 0.0168 & 0.0022 & 0.0018 & 0.0200 & \\
& & & & & & \\
\hline
\end{tabular}
\end{table}
\end{center}
\end{widetext}

\subsection{Reduced B(M1) strengths}

In principle, magnetic dipole transitions in our model are mediated
by contributions from the cluster-core relative motion and also from the
magnetic dipole moments of the cluster and core. Since the cluster
remains in its $0^+$ ground state there is no contribution from
this to the M1 transition rates. Furthermore, we intend to take
the magnetic dipole moment of the excited $2^+$ core state as zero as
well (its $0^+$ ground state is unable to make any contribution). 
This is mainly a matter of expediency since we do not have
sufficient information to make any other choice. Very few $2^+$ state
magnetic dipole moments have been measured in the $^{208}$Pb 
region \cite{[Stone]}. However, one of the few that has is for the $2^+$ state in 
$^{206}$Pb at $\sim 800$ keV which has a measured value of
$\mu < 0.03$ nm \cite{[Stone]}, which tends to support our proposed assignment of
zero. One other piece of evidence in support of a negligible $\mu(2^+)$ value
is provided by our ability to describe the magnetic dipole moment of
the lowest lying $2^+$ state of $^{224}$Ra with the simplest form of
our model. This means taking the state as a $^{210}$Pb core in its 
ground state orbited by a $^{14}$C cluster with relative orbital
angular momentum $L=2$. The magnetic dipole moment is then given by
\begin{eqnarray}
\langle \mu_J \rangle = \mu_0 
\left ( {A_1^2Z_2 + A_2^2Z_1 \over A_1A_2(A_1+A_2) } \right ) \times
\nonumber \\
\left \{ {J(J+1)+L(L+1)-I(I+1) \over 2(J+1) } \right \}
\end{eqnarray}
which for $I=0$ and $J=L=2$ reduces to
\begin{equation}
\langle \mu_2 \rangle = 2\mu_0 
\left ( {A_1^2Z_2 + A_2^2Z_1 \over A_1A_2(A_1+A_2) } \right )
\end{equation}
For $^{224}$Ra treated as $^{210}$Pb + $^{14}$C this yields a numerical
value of 0.852 nm, which compares favourably with the measured value
of $0.9 \pm 0.2$ nm \cite{[Stone]}.

Using the relative motion term alone leads to a magnetic
dipole operator in our model of
\begin{equation}
M(M1) = \sqrt{3 \over 4\pi} {A_1^2Z_2 + A_2^2Z_1 \over A_1A_2(A_1+A_2)}{\bf L}
\end{equation}
To avoid explicit use of charge and mass values we use an average value
appropriate to $^{208}$Pb + $^{14}$C, $^{20}$O, $^{24}$Ne and $^{28}$Mg of 0.42
for the charge-mass dependent factor above. Thus 
\begin{equation}
M(M1) \approx 0.42 \times \sqrt{3 \over 4\pi} {\bf L}
\end{equation}
The Weisskopf unit in these same units is $45/8\pi$ \cite{[BM_WU]}.

Table 3 shows our results for B(M1) reduced transition strengths between
states in the spectrum of Fig.1 with $J$ values differing by 1 or 0.
The values are only meant to be indicative of typical strengths,
subject to the same provisos mentioned in discussing the E2
transitions, that the level ordering might be different from
that illustrated, and accidental near-degeneracies can give
rise to strong mixing which results in fortuitously strong 
predicted transitions.
The calculated values are generally rather small, scarcely exceeding
0.01 Wu. Their possible importance lies in the existence of transitions
between the heads of the $1^+$ and $0^+$ bands, which could not
be mediated by gamma rays of any other multipolarity.

\begin{widetext}
\begin{center}
\begin{table}[htb]
\caption{Calculated cross-band reduced M1 strengths B(M1; $J_i \to J_f$)
in W.u. between $K^{\pi} = 0^+_{\rm gs}, 1^+, 0^+_{\rm ex}$ and $2^+$ bands
in heavy nuclei. See text for details.}

\begin{tabular}{lrrrrrrr}
\hline
& & & & & & & \\
Transition & $ 1^+ \to  0^+_{\rm gs}$  & 
$ 0^+_{\rm ex} \to 0^+_{\rm gs}$  & 
$ 0^+_{\rm ex} \to  1^+$  & 
$ 2^+ \to  0^+_{\rm gs}$  & 
$ 2^+ \to  1^+$  & 
$ 2^+ \to  0^+_{\rm ex}$ &
$ 0^+_{\rm ex} \to 2^+$ \\
& & & & & & & \\
\hline
$0^+ \to 1^+$ &        &        & 0.1028 &        &        &        & \\
$1^+ \to 0^+$ & 0.0084 &        &        &        &        &        & \\
$1^+ \to 2^+$ & 0.0022 &        &        &        &        &        & \\
$2^+ \to 1^+$ &        &        & 0.0045 &        & 0.0271 &        & \\
$2^+ \to 2^+$ & 0.0052 & 0.0023 & 0.0064 & 0.0001 & 0.0124 & 0.0112 & \\
$2^+ \to 3^+$ &        &        & 0.0480 &        & 0.0042 &        & \\
$3^+ \to 2^+$ & 0.0066 &        &        & 0.0001 & 0.0118 & 0.0081 & \\
$3^+ \to 3^+$ &        &        &        &        & 0.0217 &        & \\
$3^+ \to 4^+$ & 0.0067 &        &        & 0.0001 & 0.0033 & 0.0122 & \\
$4^+ \to 3^+$ &        &        & 0.0069 &        & 0.0106 &        & \\
$4^+ \to 4^+$ & 0.0112 & 0.0023 & 0.0007 & 0.0005 & 0.0111 & 0.0256 & \\
$4^+ \to 5^+$ &        &        & 0.0466 &        & 0.0055 &        & \\
$5^+ \to 4^+$ & 0.0058 &        &        & 0.0002 & 0.0050 & 0.0049 & \\
$5^+ \to 5^+$ &        &        &        &        & 0.0196 &        & \\
$5^+ \to 6^+$ & 0.0079 &        &        & 0.0002 & 0.0034 & 0.0287 & \\
$6^+ \to 5^+$ &        &        & 0.0072 &        & 0.0040 &        & \\
$6^+ \to 6^+$ & 0.0123 & 0.0018 & 0.0000 & 0.0009 & 0.0080 & 0.0267 & \\
$6^+ \to 7^+$ &        &        & 0.0508 &        & 0.0029 &        & \\
$7^+ \to 6^+$ & 0.0052 &        &        & 0.0002 & 0.0028 & 0.0014 & \\
$7^+ \to 7^+$ &        &        &        &        & 0.0158 &        & \\
$7^+ \to 8^+$ & 0.0091 &        &        & 0.0003 & 0.0031 &        & \\
$8^+ \to 7^+$ &        &        & 0.0061 &        & 0.0014 &        & 0.0371 \\
$8^+ \to 8^+$ & 0.0139 & 0.0012 & 0.0003 & 0.0012 & 0.0057 & 0.0193 & \\
$8^+ \to 9^+$ &        &        & 0.0542 &        & 0.0011 &        & \\
$9^+ \to 8^+$ & 0.0048 &        &        & 0.0003 & 0.0018 & 0.0001 & \\
$9^+ \to 9^+$ &        &        &        &        & 0.0124 &        & \\
$9^+ \to 10^+$& 0.0106 &        &        & 0.0004 & 0.0027 &        & \\
$10^+ \to 9^+$ &       &        & 0.0047 &        & 0.0005 &        & 0.0453 \\
$10^+ \to 10^+$& 0.0159& 0.0009 & 0.0005 & 0.0014 & 0.0043 & 0.0124 & \\
& & & & & & \\
\hline
\end{tabular}
\end{table}
\end{center}
\end{widetext}

\section{Comparison with existing data}

On examining the even-even Actinide nuclei for experimental B(E2) 
transitions involving any states other than those within the ground 
state band, we have found surprisingly few measurements \cite{[NNDC]}.
Only the four nuclei $^{230}$Th, $^{232}$Th, $^{234}$U and 
$^{238}$U can act as present testing grounds for our model.
We defer discussion of $^{238}$U until later becauses its structure
is so complicated, and concentrate initially on the first three 
of these nuclei. 

\begin{center}
\begin{table}[htb]
\caption{Comparison of some measured cross-band reduced 
E2 strengths B(E2; $J_i \to J_f$) for $^{230}$Th,
$^{232}$Th and $^{234}$U with possible theoretical 
equivalents from Table 2. See text for details.
Data from Refs.\cite{[NDS_230],[NDS_232],[NDS_234]}.
}

\begin{tabular}{lrrlr}
\hline
& & & & \\
$J_i(E_i {\rm keV})$ 
& $J_f(E_f {\rm keV})$ 
& B(E2)
& $J_i(K_i) \to J_f(K_f)$
& B(E2) \\ 
& & Exp & & Theo \\
& & Wu & & Wu \\
& & & & \\
\hline
& & & & \\
\multispan 5 $^{230}$Th \\
& & & & \\
\hline
& & & & \\
$2^+$(677.515) & $4^+$(174.111) & 10(4) &
$2^+(0_{\rm ex}) \to 4^+(0_{\rm gs})$ & 9.2 \\
$2^+$(677.515) & $0^+$(0.0) & 2.7(9) &
$2^+(0_{\rm ex}) \to 0^+(0_{\rm gs})$ & 0.04 \\
$2^+$(781.375) & $4^+$(174.111) & 0.37(14) &
$2^+(2) \to 4^+(0_{\rm gs})$ & 0.53 \\
$2^+$(781.375) & $2^+$(53.227) & 5.5(18) &
$2^+(2) \to 2^+(0_{\rm gs})$ & 3.0 \\
$2^+$(781.375) & $0^+$(0.0) & 2.9(9) &
$2^+(2) \to 0^+(0_{\rm gs})$ & 1.5 \\
$2^+$(1009.601) & $2^+$(677.515) & $<27$ &
$2^+(1) \to 2^+(0_{\rm ex})$ & 3.6 \\
$2^+$(1009.601) & $4^+$(174.111) & $<0.38$ &
$2^+(1) \to 4^+(0_{\rm gs})$ & 0.43 \\
$2^+$(1009.601) & $2^+$(53.227) & $<5.2$ &
$2^+(1) \to 2^+(0_{\rm gs})$ & 2.8 \\
$2^+$(1009.601) & $0^+$(0.0) & $<2.7$ &
$2^+(1) \to 0^+(0_{\rm gs})$ & 5.9 \\
& & & & \\
\hline
& & & & \\
\multispan 5 $^{232}$Th \\
& & & & \\
\hline
& & & & \\
$2^+$(774.15) & $4^+$(162.12) & $\approx 3.3$ &
$2^+(0_{\rm ex}) \to 4^+(0_{\rm gs})$ & 9.2 \\
$2^+$(774.15) & $2^+$(49.369) & $\approx 0.52$ &
$2^+(0_{\rm ex}) \to 2^+(0_{\rm gs})$ & 1.2 \\
$2^+$(774.15) & $0^+$(0.0) & 2.8(12) &
$2^+(0_{\rm ex}) \to 0^+(0_{\rm gs})$ & 0.04 \\
$2^+$(785.25) & $4^+$(162.12) & $\approx 0.13$ &
$2^+(2) \to 4^+(0_{\rm gs})$ & 0.53 \\
$2^+$(785.25) & $2^+$(49.369) & 7.2(7) &
$2^+(2) \to 2^+(0_{\rm gs})$ & 3.0 \\
$2^+$(785.25) & $0^+$(0.0) & 2.9(4) &
$2^+(2) \to 0^+(0_{\rm gs})$ & 1.5 \\
$2^+$(1387.1) & $4^+$(162.12) & 0.51(18) &
$2^+(1) \to 4^+(0_{\rm gs})$ & 0.44 \\
$4^+$(1413.8) & $3^+$(829.6) & $<12$ &
$4^+(1) \to 3^+(2)$ & 3.8 \\
$4^+$(1413.8) & $2^+$(785.25) & 23(7) &
$4^+(1) \to 2^+(2)$ & 1.9 \\
& & & & \\
\hline
& & & & \\
\multispan 5 $^{234}$U \\
& & & & \\
\hline
& & & & \\
$0^+$(809.907) & $2^+$(43.4981) & $>0.067$ &
$0^+(0_{\rm ex}) \to 2^+(0_{\rm gs})$ & 1.3 \\
$2^+$(851.74) & $4^+$(143.352) & $< 1$ &
$2^+(0_{\rm ex}) \to 4^+(0_{\rm gs})$ & 11.1 \\
$2^+$(851.74) & $2^+$(43.4981) & $<0.23$ &
$2^+(0_{\rm ex}) \to 2^+(0_{\rm gs})$ & 1.4 \\
$2^+$(851.74) & $0^+$(0.0) & $<1.3$ &
$2^+(0_{\rm ex}) \to 0^+(0_{\rm gs})$ & 0.05 \\
$2^+$(926.720) & $4^+$(143.352) & 0.28(5) &
$2^+(2) \to 4^+(0_{\rm gs})$ & 0.64 \\
$2^+$(926.720) & $2^+$(43.4981) & 4.9(8) &
$2^+(2) \to 2^+(0_{\rm gs})$ & 3.6 \\
$2^+$(926.720) & $0^+$(0.0) & 2.9(5) &
$2^+(2) \to 0^+(0_{\rm gs})$ & 1.8 \\
& & & & \\
\hline
\end{tabular}
\end{table}
\end{center}

Table 4 compares the measured cross-band B(E2) 
strengths with our calculated values, in Wu. In each case we accept
the evaluators' identifications of beta and gamma bands. We identify 
the former with our $K^{\pi}=0^+_{\rm ex}$ band and the latter with our
$K^{\pi}=2^+$ band. This allows us to compare the measured values
for transitions from these bands to the ground state band with
our corresponding calculated transition strengths.  
More interesting, is the observation
that in $^{230}$Th and $^{232}$Th there are other transitions, 
not emanating from the beta or gamma bands, which feed into the
other labelled bands, for which measured B(E2) strengths are
available. In the table we tentatively identify these gamma rays
as coming from states in our predicted $K^{\pi}=1^+$ band. We thus 
include a comparison of the strengths calculated for these putative
transitions with the otherwise unassigned experimental values.

Overall, we then obtain quite a good description of all the
cross-band transitions in these three nuclei for which measured B(E2) 
strengths are available. Agreement is generally
rather good, although we have to admit that our values for 
decays from the $2^+$ state of the beta band to the $0^+$ ground
state are generally too small (although the measured values come
with rather large uncertainties). Given the generality of our
calculation due to the uncertainties surrounding its input
parameters we feel that this is a highly satisfactory first step.

The experimental situation for $^{238}$U is much less transparent.
Lifetimes have been measured for nine separate $2^+$ states in this
nucleus. The Nuclear Data Centre 
evaluation \cite{[NDS_238]} labels these as follows: $2^+$(44.916)
is a member of the ground state band, $2^+$(966.13) and $2^+$(1037.25)
are members of two distinct beta bands, $2^+$(1060.27) is the head 
of a gamma band, and the remaining five states at excitation energies
of 1223.78, 1278.54, 1414.0, 1530.2 and 1782.3 keV are all seen in
Coulomb excitation and assigned $J^{\pi}=2^+$ through angular
distribution analysis. 

\begin{center}
\begin{table}[htb]
\caption{Comparison of some measured cross-band reduced 
E2 strengths B(E2; $J_i \to J_f$) for $^{238}$U
with possible theoretical equivalents from Table 2.
See text for details. Data from Ref.\cite{[NDS_238]}   }

\begin{tabular}{lrrlr}
\hline
& & & & \\
$J_i(E_i {\rm keV})$ 
& $J_f(E_f {\rm keV})$ 
& B(E2)
& $J_i(K_i) \to J_f(K_f)$
& B(E2) \\ 
& & Exp & & Theo \\
& & Wu & & Wu \\
& & & & \\
\hline
& & & & \\
\multispan 5 $^{238}$U \\
& & & & \\
\hline
& & & & \\
$2^+$(966.13) & $4^+$(148.38) & 3.3(14) &
$2^+(0_{\rm ex}) \to 4^+(0_{\rm gs})$ & 13.2 \\
$2^+$(966.13) & $2^+$(44.916) & 1.1(4) &
$2^+(0_{\rm ex}) \to 2^+(0_{\rm gs})$ & 1.7 \\
$2^+$(966.13) & $0^+$(0.0) & 0.38(16) &
$2^+(0_{\rm ex}) \to 0^+(0_{\rm gs})$ & 0.05 \\
$2^+$(1037.25) & $4^+$(148.38) & 2.28(23) &
$2^+(1) \to 4^+(0_{\rm gs})$ & 0.63 \\
$2^+$(1037.25) & $2^+$(44.916) & 1.23(14) &
$2^+(1) \to 2^+(0_{\rm gs})$ & 4.0 \\
$2^+$(1037.25) & $0^+$(0.0) & 1.47(16) &
$2^+(1) \to 0^+(0_{\rm gs})$ & 8.4 \\
$2^+$(1060.27) & $4^+$(148.38) & 0.33(3) &
$2^+(2) \to 4^+(0_{\rm gs})$ & 0.75 \\
$2^+$(1060.27) & $2^+$(44.916) & 5.3(4) &
$2^+(2) \to 2^+(0_{\rm gs})$ & 4.3 \\
$2^+$(1060.27) & $0^+$(0.0) & 3.04(18) &
$2^+(2) \to 0^+(0_{\rm gs})$ & 2.1 \\
$2^+$(1223.78) & $2^+$(966.13) & 32 &
& \\
$2^+$(1223.78) & $0^+$(927.21) & 27 &
& \\
$2^+$(1223.78) & $4^+$(148.38) & 0.017 &
& \\
$2^+$(1223.78) & $0^+$(0.0) & 0.29 &
& \\
$2^+$(1278.54) & $4^+$(148.38) & 0.29(3) &
& \\
$2^+$(1278.54) & $2^+$(44.916) & 0.37(5) &
& \\
$2^+$(1278.54) & $0^+$(0.0) & 0.098(9) &
& \\
$2^+$(1414.0) & $2^+$(1060.27) & 36 &
& \\
$2^+$(1414.0) & $0^+$(0.0) & 0.125 &
& \\
$2^+$(1530.2) & $2^+$(966.13) & 55 &
& \\
$2^+$(1530.2) & $4^+$(148.38) & 3.57(43) &
& \\
$2^+$(1530.2) & $0^+$(0.0) & 0.240(24) &
& \\
$2^+$(1782.3) & $2^+$(44.916) & 0.57(6) &
& \\
$2^+$(1782.3) & $0^+$(0.0) & 0.41(4) &
& \\
& & & & \\
\hline
\end{tabular}
\end{table}
\end{center}

The allocation of two separate beta bands worries us and probably
warrants further investigation. There appear to be associated
$0^+$ band heads at 927.21 and 997.23 keV for these proposed beta
bands, although this latter
state's $J^{\pi}$ value is based on an E0 transition to the ground
state. However, we also note that there is a $3^+$ state at 1105.71 keV 
(assigned to the gamma band) and a ($3^+$) state at 1059.66 keV.
The appearance of two $3^+$ states, close in energy, is a signature
of our three excited K-band model, with the two $3^+$ states belonging 
to the excited $K^{\pi}=1^+$ and $2^+$ bands. We therefore compare the 
measured reduced E2 transition strengths
to our calculated ones by accepting the evaluator's 
identification of the ground state band, the gamma band and the lower
of the two beta bands (i.e. the one based on a $0^+$ band head 
at 927.21 and containing the $2^+$(966.13) state). We treat the
decays from the $2^+$(1037.25) state as if that state were a member of
our $K^{\pi}=1^+$ band. This interpretation places the ordering of
the experimental bands as $K^{\pi} = 0^+_{\rm gs}, 0^+_{\rm ex}, 1^+$
and $2^+$. Since this differs from the ordering in our generic spectrum,
shown in Fig.1, we expect that our calculation would need 
some fine-tuning to yield a good description of $^{238}$U.

Nevertheless, Table 5 compares the experimental cross-band reduced
E2 strengths with the values calculated for our ``typical'' Actinide
nucleus. The level of agreement for transition strengths into the 
ground state band from the beta and gamma bands is good, but our 
calculated values from the proposed $1^+$ band are too large by a 
factor 3--6. This is about as good as we could expect without
fitting the model parameters specifically to $^{238}$U properties.

We leave the higher $2^+$ states in abeyance for the present time. 
Our model could accommodate (many) higher lying $2^+$ states if we 
introduced a $2^+$ cluster excitation in addition to the $2^+$ core
excitation. In fact, this cluster excitation could combine with 
the core excitation to form a total excitation ``spin'' value of 
$K_{\rm total} = 0$, 1, 2, 3 or 4 with associated K-bands running 
from K=0 to $K_{\rm total}$ in each case. Indeed, this extra 
excitation would introduce several extra beta bands. In this
context it is worth bearing in mind that a multiplicity of beta
bands can be produced by a microscopic particle-hole description
of the excitations of the system. This point has been successfully 
made by Chasman in his studies of $^{232, 234, 236}$U \cite{[Chasman]},
and his conclusions are likely to hold across all nuclei discussed in
the present paper. 

\section{Conclusion}

Consideration of a structure model where a cluster rotates around a 
core having a $0^+$ ground state and an excited $2^+$ 
state, leads to a spectrum containing a ground state $K^{\pi}=0^+$
band and three excited bands having $K^{\pi}=0^+, 1^+$ and $2^+$.
These features are observed in light nuclei such as $^{16}$O \cite{[BBR_O16]}, 
$^{24}$Mg \cite{[BMH_Mg24]} and $^{40}$Ca \cite{[Ca40]}, and there 
is no obvious reason why they should not be replicated in heavier nuclei.

We are unable to predict the energies of the excited bandheads
but are tempted to identify the excited $K^{\pi}=0^+$ and $2^+$ bands
with the widespread beta and gamma bands. We therefore choose the
free parameters of our model to place the excited bands in the vicinity
of the beta and gamma bands in Actinide nuclei. This gives the generic 
spectrum for heavy nuclei shown in Fig.1, although we accept that the 
details, and even the precise band orderings, could change with moderate
fine-tuning of the model parameters.

The $K^{\pi}=1^+$ band has not been widely seen in heavy nuclei, 
although Tables 4 and 5 tentatively indicate its presence in
$^{230}$Th, $^{232}$Th and $^{238}$U. However, our model suggests 
that it ought to be a feature common to many more heavy nuclei.
The irregularly spaced state energies emerging from our intermediate
strength coupling calculation for this band indicate that it will be 
a difficult task to identify it on the basis of $J^{\pi}$ state excitation energies 
alone. To this end, we have presented a calculation of the
reduced E2 and M1 electromagnetic transition strengths between the states
of the spectrum of Fig.1. Of course, only the
gross features and not the details should be sought in any particular nucleus,
since the B(E2) and B(M1) strengths will probably be even more
sensitive to fine-tuning of the model parameters than the spectrum.
Nevertheless, these transition rate predictions differ sufficiently from those of
the strong coupling limit of the standard collective rotational model
\cite{[Leander],[Rowe]}
that they suggest a realistic possibility of identifying the elusive $K^{\pi}=1^+$ band
and differentiating between that model and our own. 

The in-band E2 transitions of both models are very similar, and related to
each other by angular momentum coupling coefficients. The cross-band E2
transitions are very different. A combination of state mixing due to
our intermediate strength non-central interaction and the effects
of including a $2^+$ excited core lead to small but measurable 
cross-band transitions of 
typically a few Wu in our model. These transitions are absent from
the strong coupling limit of the rotational model. We also predict
small M1 transitions of typically a few hundredths of a Wu. 
The significance of the M1 transitions is mainly that they
allow the presence of gamma rays connecting the $0^+$ and
$1^+$ bandheads, where otherwise there would be nothing.

Our examination of the existing data in the Actinide nuclei is
tantalising but not conclusive. The first comment we make is that
there are surprisingly few well measured cross-band E2 transition strengths in
these nuclei. Despite this, there are some indications of the cross band E2
electromagnetic transitions predicted by our model in $^{230}$Th, 
$^{232}$Th, $^{234}$U and $^{238}$U. We have considered all reported
B(E2) strengths in these four nuclei. In three of them
there are certainly excited states, not accommodated in beta and 
gamma bands, that decay into the ground state band with strengths 
of a few Wu, in line with our model expectations. We also give
a reasonable description of cross-band E2 transitions from the
beta and gamma bands in all four nuclei. The only blemish being that the
predicted decays from the $2^+$ state of the beta band to the $0^+$ ground
state are generally too small.

In the near future we hope to apply our model in the rare-Earth region, 
where qualitatively similar results are expected. We also urge our
experimental colleagues to reexamine existing data and initiate
new experiments to improve the overall level of spectroscopy in the
Actinide region and, in particular, to verify or deny the 
existence of the proposed $K^{\pi}=1^+$ band which is a key
distinguishing characteristic of our model.

\vskip 1truecm
\noindent
{\bf Acknowledgements:} \ \
S.M.P. would like to thank  the National Research Foundation of S.A. 
and the University of Cape Town, for financial support.

\begin{figure*}
\includegraphics{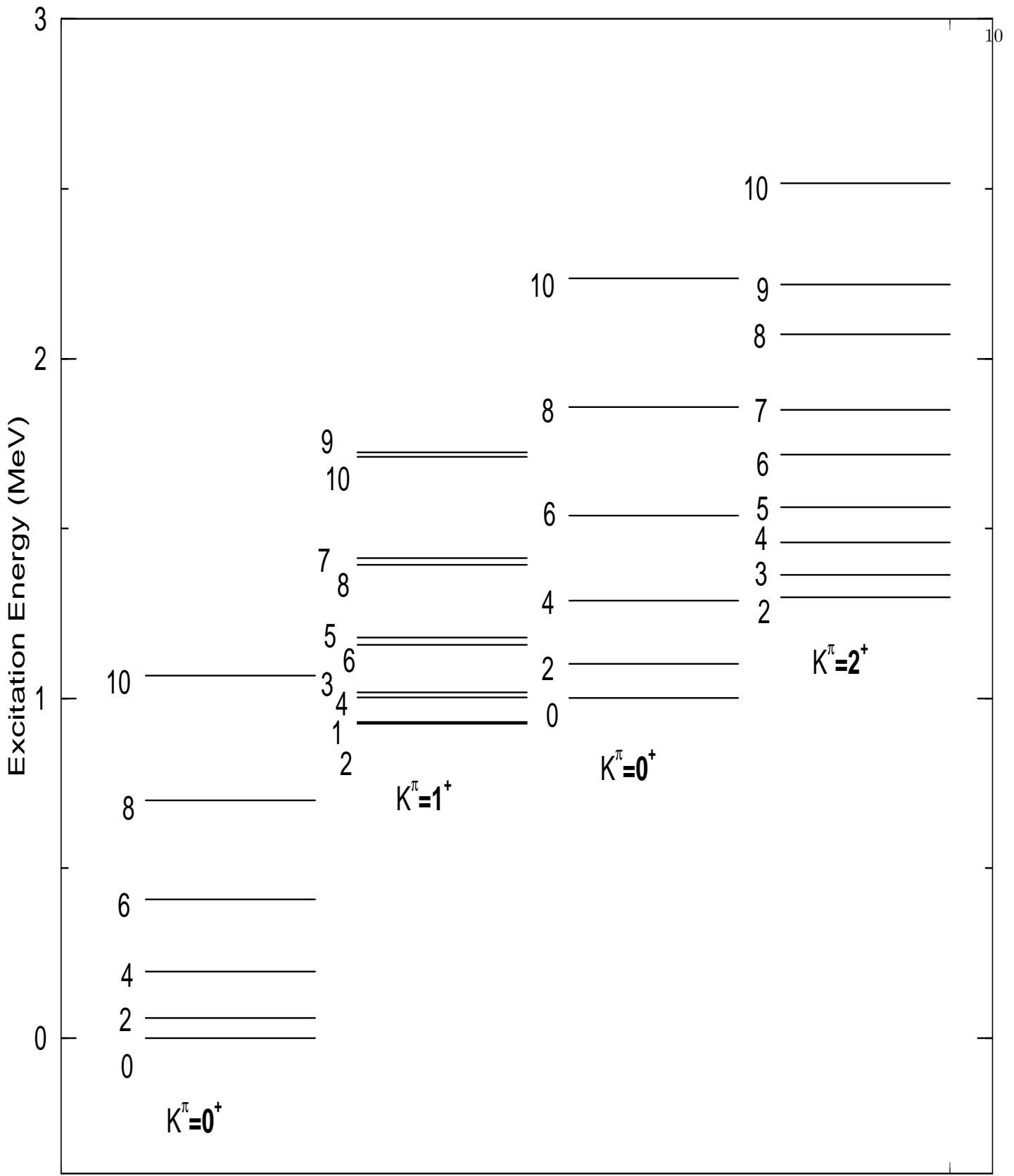}
\caption{\label{fig1:wide}
Generic positive parity spectrum obtained by coupling a $0^+$
ground state and an excited $2^+$ state to a rotor. We
use a rotational parameter $\alpha = 0.01$ MeV, a core
excitation energy $\epsilon = 0.8$ MeV and a radial integral
parameter $\beta = -0.178$ MeV, exactly as in Ref.\cite{[BMP_Spectrum]}
}
\end{figure*}

\end{document}